\documentclass[preprint,aps,prb,superscriptaddress,floatfix]{revtex4-1}
\usepackage{graphicx}
\usepackage{amsmath}
\usepackage{hyperref}
\usepackage{bm}
\renewcommand{\vec}[1]{\bm{#1}}

\begin{document}

\title{Rare-earth/transition-metal magnetic interactions in pristine and (Ni,Fe)-doped 
YCo$_5$ and GdCo$_5$}
\author{Christopher E. Patrick}
\author{Santosh Kumar}
\author{Geetha Balakrishnan}
\author{Rachel S. Edwards}
\author{Martin R. Lees}
\author{Eduardo Mendive-Tapia}
\affiliation{Department of Physics, University of Warwick, Coventry CV4 7AL, United Kingdom}
\author{Leon Petit}
\affiliation{Daresbury Laboratory, Daresbury, Warrington WA4 4AD, United Kingdom}
\author{Julie B. Staunton}
\email[]{j.b.staunton@warwick.ac.uk}
\affiliation{Department of Physics, University of Warwick, Coventry CV4 7AL, United Kingdom}
\date{\today}

\begin{abstract}
We present an investigation into the intrinsic magnetic properties
of the compounds YCo$_5$ and GdCo$_5$, members of the RETM$_5$ class of permanent
magnets (RE = rare earth, TM = transition metal).
Focusing on Y and Gd 
provides direct insight into both the TM magnetization and RE-TM
interactions without the complication of strong crystal field effects.
We synthesize single crystals of YCo$_5$ and GdCo$_5$ using the 
optical floating zone technique and measure the magnetization 
from liquid helium temperatures up to 800~K.
These measurements are interpreted through calculations based
on a Green's function formulation of density-functional
theory, treating the thermal disorder of the local magnetic moments within
the coherent potential approximation.
The rise in the magnetization of GdCo$_5$ with temperature is shown to arise
from a faster disordering of the Gd magnetic moments compared to the 
antiferromagnetically aligned Co sublattice.
We use the calculations to analyze the different Curie temperatures of
the compounds and also compare the molecular (Weiss) fields at the RE site with
previously published neutron scattering experiments.
To gain further insight into the RE-TM interactions, we perform substitutional
doping on the TM site, studying the compounds RECo$_{4.5}$Ni$_{0.5}$, RECo$_4$Ni
and RECo$_{4.5}$Fe$_{0.5}$.
Both our calculations and experiments on powdered samples find
an increased/decreased magnetization with Fe/Ni-doping
respectively.
The calculations further reveal a pronounced dependence 
on the location of the dopant atoms
of both the Curie temperatures and the 
Weiss field at the RE site.
\end{abstract}

\maketitle

\section{Introduction}

The discovery of the
favorable magnetic properties of SmCo$_5$ fifty years ago\cite{Strnat1967} 
triggered a technological revolution
based on rare-earth transition-metal (RE-TM) permanent magnets.\cite{Coey2011}
In SmCo$_5$, the strong magnetism of Co combines with the
large magnetocrystalline anisotropy of localized Sm-4$f$ electrons
to form an excellent permanent magnet.
As well as having provided the blueprint for the development of the now
ubiquitous Nd-Fe-B RE-TM magnet class,\cite{Buschow1997}
Sm-Co compounds still play an important role in
commercial applications due to their superior high-temperature 
performance.\cite{Gutfleisch2011}
SmCo$_5$ also remains interesting from a fundamental viewpoint, 
since understanding precisely how the 
complicated interplay of localized and delocalized electrons
affects the anisotropy and magnetization 
is a significant challenge for electronic structure theory.\cite{Richter1998}

\begin{figure}
\centering
\includegraphics{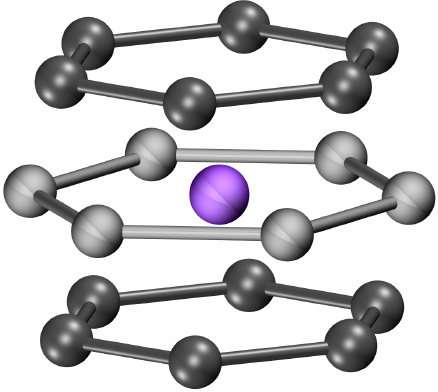}
\caption{
Ball-and-stick representation of the RETM$_5$ 
crystal structure, showing the RE site (purple)
and two TM sublattices: TM$_{2c}$ (light gray, in
plane with the RE) and TM$_{3g}$ (dark gray)
where $2c$ and $3g$ refer to the Wyckoff positions.
\label{fig.ballstick}
}
\end{figure}

SmCo$_5$ belongs to the RETM$_5$ family of
permanent magnets which crystallize in the CaCu$_5$ structure
($P6/mmm$) whose unit cell is formed of alternating RETM$_{2c}$/TM$_{3g}$
layers (Fig.~\ref{fig.ballstick}).\cite{Kumar1988}
This relatively simple crystal structure, paired with
the diverse behavior exhibited by magnets with different
RE,\cite{Ermolenko1976} make the RETM$_5$ family an
appealing playground for the investigation of RE-TM
interactions.
In particular, a hierarchy of complexity can be established
beginning with RE = Y (i.e.\ a nonmagnetic RE with no 4$f$
electrons), followed by RE = Gd (a half-filled 4$f$ shell
whose spherical symmetry removes a number of complications
involving the spin-orbit interaction and crystal
field [CF]), and finally a generic RE with a partly-filled $4f$
shell, like Sm.
The different energy scales involved in the interactions\cite{Kuzmin2008}
allow this hierarchical approach to yield a quite
general understanding of the TM-TM, RE-TM and RE-CF
interactions respectively
(the label ``RE-CF'' used in this sense
denotes the interactions of the non-spherical 4$f$ charge
cloud with the crystal field).
An early example of this approach is the empirical 
subtraction of the magnetization curve of YCo$_5$ from other RECo$_5$
compounds in order to observe the RE magnetism.\cite{Yermolenko1980}

In order to lay the essential groundwork for
the future study of compounds where RE-CF 
interactions are also important, here
we concentrate on YCo$_5$ and GdCo$_5$.
Our strategy is to synthesize and characterize samples and
then interpret the results using first-principles calculations
based on density-functional theory (DFT).
In particular, by applying the disordered local moment (DLM) 
picture\cite{Gyorffy1985} we aim to understand the evolution of magnetic properties
with temperature, an aspect which is of obvious practical importance.
To this end we have grown single crystals of YCo$_5$ and GdCo$_5$ 
using the optical floating zone technique (FZT)
and measured the evolution of the magnetization up to 800~K.
Our DFT-DLM calculations are able to explain
both the contrasting temperature dependences of the two compounds 
and also the experimentally-observed higher Curie
temperature of GdCo$_5$.
To further elucidate the RE-TM physics underlying
these and other permanent magnets, we have also synthesized 
polycrystalline samples where Co was substituted with Fe (Ni),
which show an increase (decrease) in Curie
temperature and magnetization. 
Our calculations reproduce this behavior, and further explore 
the dependence of these properties on the 
crystallographic site occupied by the dopants.
Indeed, the calculations find an unusual ferromagnetic
RE-TM interaction between Gd and Fe when the atoms
occupy nearest neighbor sites.

The rest of this manuscript is organized as follows.
In section~\ref{sec.methods} we describe the experimental
and computational approach used in our study.
In section~\ref{sec.results} we report our findings,
beginning with pristine YCo$_5$ and GdCo$_5$ (section~\ref{sec.YCo5GdCo5})
and extending to the doped samples (section~\ref{sec.firstdope} onwards).
In section~\ref{sec.conclusions} we summarize our results and present our conclusions.

\section{Experimental and theoretical approach}

\label{sec.methods}
\subsection{Experimental overview}
\label{sec.expoverview}

Owing to its technological importance
the RECo$_5$ family has been
the subject of extensive investigation
for several decades, with experiments
investigating the temperature dependence 
of magnetization and anisotropy of pristine
RECo$_5$ compounds.\cite{Franse1993, Nassau1960, 
Nesbitt1962, Burzo1972, 
Okamoto1973, Katayama1976,Uehara1982, Ermolenko1979, Yermolenko1980,
Radwanski1986,Ballou1986,
Radwanski1992, Gerard1992, Kuzmin2004,
Frederick1974,
Klein1975,Tatsumoto1971,
Alameda1981, Schweizer1980, Yamaguchi1983}
However,  the growth of single crystals remains challenging~\cite{Miller1973,Katayama1974}
and to the best of our knowledge our study represents the first successful attempt
to grow single crystals of RECo$_5$ compounds using the optical FZT.
Furthermore, while there are a number of studies
investigating specific cases of TM-doped RECo$_5$ compounds,\cite{
Buschow1976, Ermolenko1977, Deportes1976, Chuang1982, Crisan1995, 
Laforest1973, Rothwarf1973, Taylor1975, Paoluzi1994, Drzazga1989, 
Liu1991} our study
tackles both Ni and Fe-doping 
on both YCo$_5$ and GdCo$_5$.
By synthesizing all compounds under
the same experimental protocols, 
we can more rigorously compare trends measured across
the series to our calculations.

\subsection{Experimental approach}
Polycrystalline samples of RECo$_{5-x}$Ni$_x$  (RE = Y, Gd, $x$ = 0, 0.5, 1.0)
and YCo$_{4.5}$Fe$_{0.5}$ were synthesized by arc melting the constituent
elements in the appropriate proportions on a water-cooled copper crucible
in an argon atmosphere.
The ingots were melted, flipped and remelted to ensure homogeneity. 
No significant changes in weight were observed after melting.
Structural characterization was performed by recording powder x-ray
diffraction (XRD) patterns of the as-cast samples using a 
Panalytical Empyrean x-ray diffractometer with a Co target.
Single crystals of YCo$_5$ and GdCo$_5$ were grown using a four-mirror
Xenon arc lamp optical image furnace 
(CSI FZ-T-12000-X\_VI-VP, Crystal Systems Inc., Japan) using 
the floating zone technique.
The polycrystalline rods for the crystal growth were synthesized by
arc melting.
The single crystals obtained were aligned using a backscattered X-ray 
Laue system (Photonic-Science Laue camera).
Platelet-shaped crystal samples with the crystallographic $c$ axis normal 
to the plane of the plates were obtained from the as-grown crystal boules.
The measured lattice constants are reported in Appendix~\ref{app.struc}.

Magnetization measurements were carried out using a Quantum Design Magnetic 
Property Measurement System (MPMS) superconducting quantum interference 
device (SQUID) magnetometer.
An oven option was used for measurements between 400 and 800~K.
Magnetization measurements on the single crystals were performed with the 
applied magnetic field along the easy axis of magnetization
so as to obtain the saturated moment values.
Below 400~K the data were collected at intervals of 10~K, while above 
400~K the data were recorded while warming at 10~K/minute.
In the case of the doped polycrystalline samples, the magnetization versus field 
curves were recorded using powder samples, with the grains free to rotate under 
the influence of the magnetic field, so as obtain a best estimate of the 
saturated magnetic moments.

\subsection{Theoretical overview}

Following on from theoretical studies of RECo$_5$ compounds based
on experimentally-parameterized CF-models,\cite{Burzo1972,Ermolenko1979,Yermolenko1980,
Radwanski1986, Ballou1986, Franse1988, Radwanski1992,Gerard1992,
Klein1975,Alameda1981,Crisan1995,Liu1991, Tiesong1991,Zhao1991}
first-principles investigations
became possible thanks to developments in
density-functional theory.\cite{Kueblerbook,Richter1998,Skomski2013}
A greater number of first-principles studies of YCo$_5$\cite{Nordstrom1992,
Daalderop1996,Yamada1999,Steinbeck20012, Uebayashi2002, Kashyap2003,Ishikawa2003,
Larson2003, Larson2004,Rosner2006,Koudela2008,
Benea2010, Liu2010,Matsumoto2014,Kumar20142, Ochi2015}
can be found compared to GdCo$_5$,\cite{Richter1998,Beuerle1994,
Liu1994,Kuzmin2004} presumably due to
the difficulty of finding an approximate
exchange-correlation functional capable of describing
the localized Gd-$4f$ electrons in DFT.
However, most of these studies were performed 
in a conventional wavefunction-based
framework, which is best suited to describing
pristine systems at zero temperature.
Although dopants can be modeled within this framework 
via calculations on supercells\cite{Steinbeck20012,
Uebayashi2002,Larson2003,Larson2004,Liu2010}
or by using virtual atoms,\cite{Yamada1999,Ishikawa2003}
the former approach quickly becomes costly in terms
of size convergence
while the latter cannot capture the full chemistry
of the problem.
Meanwhile the calculation of finite-temperature properties
in a wavefunction-based framework is generally limited to 
obtaining critical temperatures based on an assumed Heisenberg
model and pairwise interactions.\cite{Kashyap2003,Liu2010}

Here, instead of wavefunctions we use the 
Korringa-Kohn-Rostocker multiple-scattering formulation
of DFT\cite{Gyorffy1979} combined with the coherent potential 
approximation (KKR-CPA)\cite{Ebert2011} and 
the disordered local moment picture,\cite{Gyorffy1985}
which reformulates the problem of compositional and thermal magnetic disorder 
in terms of impurity scattering.
Ref.~\citenum{Benea2010} used this approach to study
the zero-temperature properties of (Al,Si)-doped YCo$_5$,
while Ref.~\citenum{Matsumoto2014} investigated the finite
temperature properties of pristine YCo$_5$.
The current study combines the computational machinery of the KKR-CPA,
the DLM picture, and the local self-interaction correction 
developed in Ref.~\citenum{Lueders2005}
to tackle the full problem of
the temperature-dependent  magnetic properties 
of pristine and transition-metal-doped YCo$_5$ and GdCo$_5$.

\subsection{Theoretical approach}
\label{sec.theory}

We follow closely the computational approach described in Ref.~\citenum{Matsumoto2014}
and refer the reader to that and other works\cite{Gyorffy1985,Staunton2006,Staunton2014,Khan2016} 
for a detailed presentation of the underlying theory.
Here we define and describe the key quantities used in our analysis.
The technical details of our calculation are reported in Appendix~\ref{app.compdet}.

The key concept in the DLM picture is the assignment of 
a local magnetic moment $\vec{\mu_i}$
to each magnetic ion, which we label by the subscript $i$.
This local moment undergoes fluctuations on the timescale associated with
spin-wave excitations, but is stable over the much shorter timescale
associated with electron motion.\cite{Gyorffy1985}
Introducing the unit vectors $\vec{\hat{e}_i} = \vec{\mu_i}/\mu_i$
to denote the orientations of the local moments, a  ``good'' local moment
system is one where the magnitudes $\{\mu_i\}$ do not depend strongly on
the orientations $\{\vec{\hat{e}_i}\}$.\cite{Staunton2014}
The statistical mechanics of such a system is determined by the
thermodynamic potential $\Omega(\{\vec{\hat{e}_i}\})$ which in principle
could be obtained from finite-temperature constrained DFT on a large
supercell containing many local moments.\cite{Gyorffy1985}
However,  the number of such calculations required to adequately sample
the large configurational space spanned by all of the possible 
orientations $\{\vec{\hat{e}_i}\}$ makes such a direct approach intractable.

To proceed, we instead approximate the statistical mechanics of the local moments
with that of an auxiliary system,
defined in terms of a model potential
\begin{equation}
\Omega_0(\{\vec{\hat{e}_i}\}) =  -\sum_i \vec{h_i}\cdot \vec{\hat{e}_i}.
\label{eq.omega0}
\end{equation}
The vectors $\{\vec{h_i}\}$ are parameters of the model with
units of energy; they play the role of molecular fields experienced by
the local moments, and we refer to them as ``Weiss fields''.
Although not written explicitly,  the Weiss fields depend on
temperature.
The number of distinct Weiss fields can be chosen to equal the number of 
crystallographically-distinct
sites in the unit cell; however, we emphasize that the sum in equation~\ref{eq.omega0}
is over all of the local moments, distributed over the entire crystal.

The potential of equation~\ref{eq.omega0} yields a probability distribution
for observing a set of local moment orientations $\{\vec{\hat{e}_i}\}$
as 
\begin{equation}
P_0(\{\vec{\hat{e}_i}\})=  \prod_i \frac{1}{Z_{0i}}\exp[\vec{\lambda_i}\cdot \vec{\hat{e}_i}]
\label{eq.P0}
\end{equation}
with $Z_{0i} = (4\pi/\lambda_i)\sinh(\lambda_i)$,
and we have introduced the dimensionless quantities 
$\vec{\lambda_i} = \beta \vec{h_i}$ (where $1/\beta = k_B T$).
The thermal averages of certain quantities with respect to the model probability 
distribution $P_0$ can be performed analytically,
e.g.\ the thermally-averaged orientation of a local moment
$\vec{m_i}(T) =  \langle \vec{\hat{e}_i} \rangle_{0,T}$:
\begin{eqnarray}
\vec{m_i}(T)
&=& 
\int d\vec{\hat{e}_i} 
\vec{\hat{e}_i}
\frac{\exp[\vec{\lambda_i}\cdot \vec{\hat{e}_i}]}
{Z_{0i}} 
\prod_{j\neq i}
\int d\vec{\hat{e}_j} 
\frac{\exp[\vec{\lambda_j}\cdot \vec{\hat{e}_j}]} 
{Z_{0j}}\nonumber \\
&=&
\vec{\hat{\lambda}_i} L(\lambda_i) \label{eq.m_i}
\end{eqnarray}
with $L(\lambda_i) = \coth(\lambda_i) - 1/\lambda_i$.
$\vec{m_i}(T)$ serve as local order parameters which vanish
above the Curie temperature.
The integrations are over the angular variables $(\theta_i,\phi_i)$
where $\vec{\hat{e}_i} = (\sin\theta_i\cos\phi_i,\sin\theta_i\sin\phi_i,\cos\theta_i)$.

The link between the model parameters $\{\vec{h_i}\}$
and the exact potential $\Omega(\{\vec{\hat{e}_i}\})$
is established through use of the thermodynamic inequality\cite{Gyorffy1985}
\begin{equation}
F(T) \leq F_0(T) - \langle \Omega_0 \rangle_{0,T}
+ \langle \Omega \rangle_{0,T}.
\label{eq.inequal}
\end{equation}
Here $F$ is the exact, unknown free energy. while $F_0$ is
the free energy calculated with the model potential (an analytical
function of the Weiss fields).
The thermal averages $\langle \rangle_{0,T}$ of the exact and model potentials are
calculated with respect to the model probability distribution,
emphasized by the $0$ subscript.
We define the optimal Weiss fields to be those
which minimize the right hand side of equation~\ref{eq.inequal}.
Performing the minimization yields
\begin{equation}
\vec{h_i} = -\nabla_{m_i} \langle \Omega \rangle_{0,T}
\label{eq.hdiff}
\end{equation}
which can be equivalently written as an integral expression,\cite{Gyorffy1985}
\begin{equation}
\vec{h_i} = -\frac{3}{4\pi} \int d\vec{\hat{e}_i} 
\langle \Omega \rangle_{0,T}^{\vec{\hat{e}_i} }
\
\vec{\hat{e}_i}
\label{eq.hint}
\end{equation}
where $\langle \rangle_{0,T}^{\vec{\hat{e}_i} }$
denotes a partial thermal average, i.e.\ the appropriately-weighted 
integration over all
local moment orientations except $\vec{\hat{e}_i}$.

Equation~\ref{eq.hint} is the expression used to evaluate
the Weiss fields within the KKR-CPA formalism.
One can draw the analogy with the simulation of alloys,
where the local moment disorder determined by the probability
$\exp[\vec{\lambda_j}\cdot \vec{\hat{e}_j}]$
is replaced with compositional disorder determined by a probability
(concentration) $c_X$.
The CPA was originally developed with the alloy problem in mind,\cite{Gyorffy1979}
and its extension to magnetic systems still retains the
possibility of including such compositional disorder.
Therefore for a given set of $\{\vec{\lambda_i}\}$ and concentrations,
one can evaluate the Weiss fields subject to the local spin density
and coherent potential approximations.
More details on the scattering theory underlying the evaluation of 
equation~\ref{eq.hint} are given e.g.\ in Ref.~\citenum{Matsumoto2014}.

Since the Weiss fields themselves determine the probability distribution used 
in the partial thermal average, equation~\ref{eq.hint} must be solved
self-consistently.
Indeed the critical (Curie) temperature $T_\mathrm{C}$ for the onset of magnetic
order is the highest temperature
at which such self-consistent solutions 
can be found.
Once the Weiss fields have been determined at a particular temperature,
the model probability distribution $P_0$ can be fed into additional
KKR-CPA calculations to calculate
thermal averages of spin and orbital moments 
(and in principle other quantities such as the torque)\cite{Staunton2006}
as $\langle A \rangle_{0,T}$, where $A$
is the appropriate quantum mechanical operator.

\section{Results and Discussion}
\label{sec.results}
\subsection{Magnetization vs.\ temperature of pristine YCo$_5$ and GdCo$_5$ }
\label{sec.YCo5GdCo5}

\begin{figure*}
\includegraphics{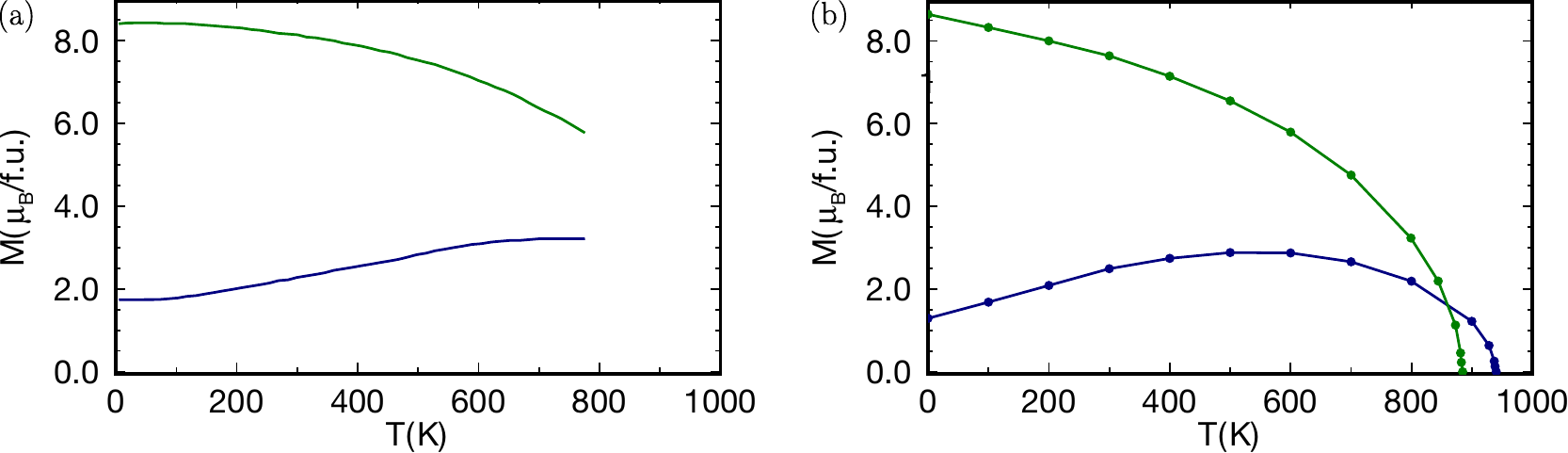}
\caption{Magnetization vs temperature (a) measured on single-crystal samples
and (b) calculated in the DLM picture, for YCo$_5$ (green)
and GdCo$_5$ (blue).
The straight lines connecting points in (b) are guides to the eye.
\label{fig.MvTpristine}
}
\end{figure*}

We begin our study with pristine YCo$_5$ and GdCo$_5$.
The experimentally-determined magnetization vs.\ temperature ($M$v$T$) curves
are shown in Fig.~\ref{fig.MvTpristine}(a).
These data were measured for our single crystals in an 
applied field of 1 or 2~T directed along the easy $c$-axis
for temperatures below and above 400~K, respectively.
As we discuss in Sec.~\ref{sec.MvH}, this field is sufficient
to saturate the magnetization.
We see from Fig.~\ref{fig.MvTpristine}(a) that YCo$_5$ 
behaves like an ordinary ferromagnet, with a monotonic
decrease in magnetization per formula unit from
8.41$\mu_B$ at 4~K to 6.38$\mu_B$ at 700~K.
The magnetization of GdCo$_5$ meanwhile increases, 
from 1.74$\mu_B$ at 4~K to 3.21$\mu_B$ at 700~K.

For the behavior of the magnetization at higher temperatures we
refer to previously-reported measurements,\cite{Nassau1960,Okamoto1973,Yermolenko1980} 
which show the magnetization of GdCo$_5$ to start decreasing at temperatures in the 
region of 700--800~K
(a lower value of 600~K was found in Ref.~\citenum{Burzo1972}).
The reported Curie temperatures\cite{Nassau1960,Burzo1972,Okamoto1973,Yermolenko1980,Franse1993} 
for GdCo$_5$ fall in the range 1000--1030~K 
compared to the lower range
of 980--1000~K~\cite{Tatsumoto1971,Klein1975, Yermolenko1980} reported for YCo$_5$.
The review article of Ref.~\citenum{Buschow1977} gives values of 
1014~K and 987~K for the $T_\mathrm{C}$ of GdCo$_5$ and YCo$_5$ respectively.
 
\begin{table}
\begin{ruledtabular}
\begin{tabular}{lll}
  & YCo$_5$ & GdCo$_5$ \\
\hline
$\mu_\mathrm{RE}$          & ---      &-7.32/-0.01 \\
$\mu_{\mathrm{Co}_{2c}}$   &1.62/0.15 & 1.57/0.15  \\
$\mu_{\mathrm{Co}_{3g}}$   &1.64/0.06 & 1.67/0.05  \\
$\mu_{\mathrm{Co}_{3g'}}$ &1.63/0.08 & 1.65/0.07   \\
$\mu_{\mathrm{Tot,calc}}$ & 8.64 & 1.29 \\
$\mu_{\mathrm{Tot,exp}}$ & 8.41\footnotemark[1] & 1.74\footnotemark[1]\\
$\mu_{\mathrm{Tot,exp}}$ & 
8.3\footnotemark[2]
8.3\footnotemark[3]
8.13\footnotemark[4]
7.9\footnotemark[5]
& 
1.55\footnotemark[3]
1.72\footnotemark[6]
1.68\footnotemark[7]
1.42\footnotemark[8]
\footnotetext[1]{Current work, 4~K, optical FZT}
\footnotetext[2]{Ref.~\citenum{Klein1975}, 0~K, r.f.\ melting + heat treatment + grinding}
\footnotetext[3]{Ref.~\citenum{Yermolenko1980}, 4~K, r.f.\ melting + heat treatment + grinding}
\footnotetext[4]{Ref.~\citenum{Frederick1974}, 0~K, induction zone melting + grinding}
\footnotetext[5]{Ref.~\citenum{Tatsumoto1971}, 0~K, plasma jet melting + heat treatment + grinding}
\footnotetext[6]{Ref.~\citenum{Kuzmin2004}, 5~K, r.f.\ melting + heat treatment + grinding}
\footnotetext[7]{Ref.~\citenum{Uehara1982}, 12~K, arc melting + grinding}
\footnotetext[8]{Ref.~\citenum{Okamoto1973}, 0~K, plasma jet melting + heat treatment + grinding}
\end{tabular}
\end{ruledtabular}
\caption{Magnetic moments in $\mu_B$ (per atom or formula unit) 
for pristine YCo$_5$ and GdCo$_5$.
The calculations were performed at 0~K for magnetization along
the [101] direction.
The experimental values have been measured by us or reported
previously in the literature; note the 0~K values were obtained
by extrapolation.
The calculations have been resolved into
spin/orbital contributions with
the Co atoms labeled as in Fig.~\ref{fig.ballstick};
note that the magnetization breaks the symmetry of the $3g$ sublattice,
giving rise to a distinct contribution ($3g'$) from the Co atom
at the $0 \frac{1}{2} \frac{1}{2}$ position.
\label{tab.zeroKM}}
\end{table}
Our calculated $M$v$T$ curves for YCo$_5$ and GdCo$_5$ are shown in 
Fig.~\ref{fig.MvTpristine}(b).
Pleasingly, we see the same contrasting behavior 
between the compounds as observed experimentally.
Our calculated $T_\mathrm{C}$ values are
885 and 940~K for YCo$_5$ and GdCo$_5$ respectively,
while the 0~K magnetizations are calculated to be 8.64$\mu_B$
and 1.29$\mu_B$.
Table~\ref{tab.zeroKM} gives the 
decomposition of the magnetization into local spin and orbital moment 
contributions.
As shown in Table~\ref{tab.zeroKM} and as realized from early
experiments,\cite{Nesbitt1962} the RE and TM sublattices align 
antiferromagnetically, accounting for the $\sim$7$\mu_B$ difference
between YCo$_5$ and GdCo$_5$.

\subsection{Comparison of calculations and experiment}
Table~\ref{tab.zeroKM} also lists magnetizations
measured by us and reported in previous literature on single crystals.
It is apparent that the calculations find a larger magnetization 
for YCo$_5$  and smaller one for GdCo$_5$ than measured experimentally.
However,  the size of the discrepancy (0.4$\mu_B$) is of 
the same magnitude as the change in magnetization
on applying an empirical orbital polarization correction
(0.5--0.8$\mu_B$/f.u.\cite{Nordstrom1992,Daalderop1996,Steinbeck20012}),
the size of the induced moment on Y ($\sim$0.3$\mu_B$,\cite{Daalderop1996,Steinbeck20012}
which we disregard) and the variation of the magnetization depending
on the choice of spherical approximation
for the potential ($\sim$0.2~$\mu_B$).\cite{Daalderop1996}
Therefore we find the current level of agreement between calculated and
experimental magnetizations to be acceptable.
Comparing our experimental magnetizations to previously-reported values
we find our values to lie in at the higher end of the range.
However,  as emphasized by Table~\ref{tab.zeroKM} 
our study is unique using the optical FZT to synthesize 
the samples.

Regarding $T_\mathrm{C}$, the calculations reproduce the experimental ordering
of YCo$_5$ and GdCo$_5$ but the calculated values are smaller than the 
experimentally-reported ones by approximately 100~K.
Usually one would expect an overestimate of $T_\mathrm{C}$ in a mean-field
approach.
A possible reason for this discrepancy is the use of the atomic-sphere 
approximation (ASA) to describe the potential (App.~\ref{app.compdet}).
We note that using a more severe muffin tin approximation 
further reduces the values of $T_\mathrm{C}$ to 774~and~749~K, so conversely
a calculation using a more accurate potential might be expected 
to yield increased values of $T_\mathrm{C}$.
Unfortunately such full-potential calculations are not yet 
feasible within our computational framework.

\begin{figure}
\includegraphics{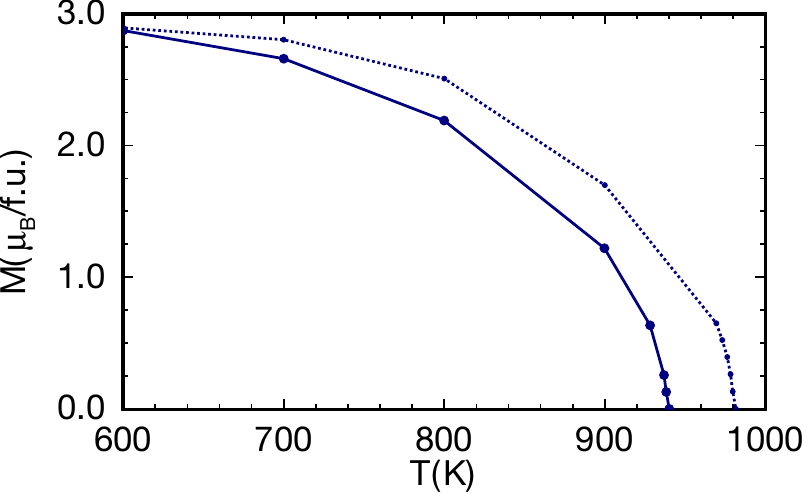}
\caption{
Magnetization calculated for GdCo$_5$ using the 300~K lattice
parameters (solid line, larger circles; cf.\ Fig.~\ref{fig.MvTpristine})
or using the temperature-dependent lattice parameters
reported in Ref.~\citenum{Andreev1991} (dotted line, smaller circles).
Note that the temperature-dependent lattice data points  $>950$~K were all calculated
using the same lattice parameters, measured at 1000~K in Ref.~\citenum{Andreev1991}.
\label{fig.GdCo5temp}
}
\end{figure}
An interesting additional consideration is the role of magnetostructural 
interactions.
The data in Fig.~\ref{fig.MvTpristine}(b) were calculated
using the lattice constants measured at 300~K as reported
in Refs.~\citenum{Andreev1991}~and~\citenum{AndreevHMM}, namely
$a$, $c$ = 4.979, 3.972~\AA \ for GdCo$_5$ and
$a$, $c$ = 4.950, 3.986~\AA \ for YCo$_5$.
For GdCo$_5$ we have investigated the effect of lattice thermal
expansion, by recalculating the magnetization at temperatures $>600$~K
using the lattice parameter data given in Ref.~\citenum{Andreev1991}.
The comparison of magnetizations obtained for the fixed or expanding 
lattices are shown in Fig.~\ref{fig.GdCo5temp}.
When lattice expansion is taken into account, the calculated $T_\mathrm{C}$ 
increases by 42~K to 982~K.
The sensitivity of magnetic coupling to the lattice parameters is explored further
in Sec.~\ref{sec.orderparam}.

As a general note, we see that in the $T\rightarrow0$ limit, the gradients 
of the experimental $M$v$T$ curves go to zero whilst those of the 
calculated curves do not.
This behavior is a simple consequence of us using a classical rather than quantized  
expression to describe the statistical mechanics of the local
moments (equation~\ref{eq.omega0}).

\subsection{The disordering of Gd in GdCo$_5$}
\begin{figure}
\includegraphics{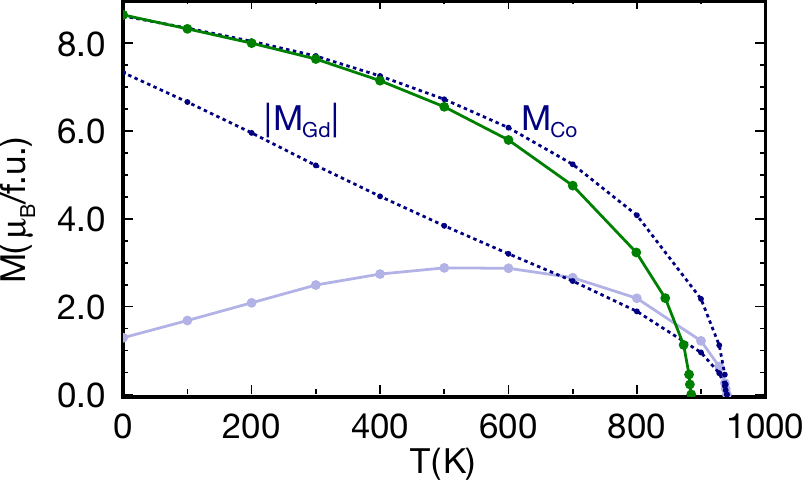}
\caption{
Decomposition of the magnetization of GdCo$_5$ (faint solid blue line) 
into contributions from the Gd and Co sublattices $M_\mathrm{Gd}$ and
$M_\mathrm{Co}$ (dotted lines, small circles).
Note that the sublattice magnetizations point antiparallel, so the resultant
magnetization is $M_\mathrm{Co} - | M_\mathrm{Gd}|$.
The calculated magnetization of YCo$_5$ (green solid line) is also
shown for comparison.
\label{fig.GdCo5resolved}
}
\end{figure}

In order to better understand the temperature evolution of
the magnetism in GdCo$_5$ it is instructive to decompose
the total magnetization into contributions from the antialigned
Gd and Co sublattices, as shown in Fig.~\ref{fig.GdCo5resolved}.
First we note that below 400~K, the Co contribution $M_\mathrm{Co}$
is indistinguishable from the $M$v$T$ curve of YCo$_5$,
showing that replacing Y with Gd (i.e.\ moving from 
a nonmagnetic to magnetic RE) has a negligible effect
on the TM ordering.
This observation is in agreement with the established
hierarchy of interaction strengths in RE-TM magnets\cite{Franse1993}
and justifies the practice of subtracting the YCo$_5$ curve
from RECo$_5$ measurements to observe the RE contribution
cited in the Introduction.\cite{Yermolenko1980}
However,  as discussed in Sec.~\ref{sec.orderparam} the
RE does have a noticeable effect on the TM sublattice
at higher temperatures.

Now considering the Gd contribution, we see the magnitude
of the magnetization $|M_\mathrm{Gd}|$ decreases more
quickly with temperature than $M_\mathrm{Co}$.
As a result the total magnetization $M_\mathrm{Co} - |M_\mathrm{Gd}|$
increases with temperature.
As shown in Fig.~\ref{fig.GdCo5resolved} the decrease
in $|M_\mathrm{Gd}|$ is effectively linear up to temperatures
of 800~K, while $M_\mathrm{Co}$ displays Brillouin function
behavior.
Consequently there is a temperature ($\sim$600~K)
where the gradients of $M_\mathrm{Co}$ and $|M_\mathrm{Gd}|$
are equal, corresponding to a peak in the total magnetization,
before $M_\mathrm{Co}$ undergoes a faster decrease close to $T_\mathrm{C}$.
In Sec.~\ref{sec.WeissGd} we reexamine this behavior in terms
of the Weiss field at the RE site and compare to low-temperature
experimental data.

\subsection{Order parameter expansion of $\Omega_0$}
\label{sec.orderparam}

\begin{figure}
\includegraphics{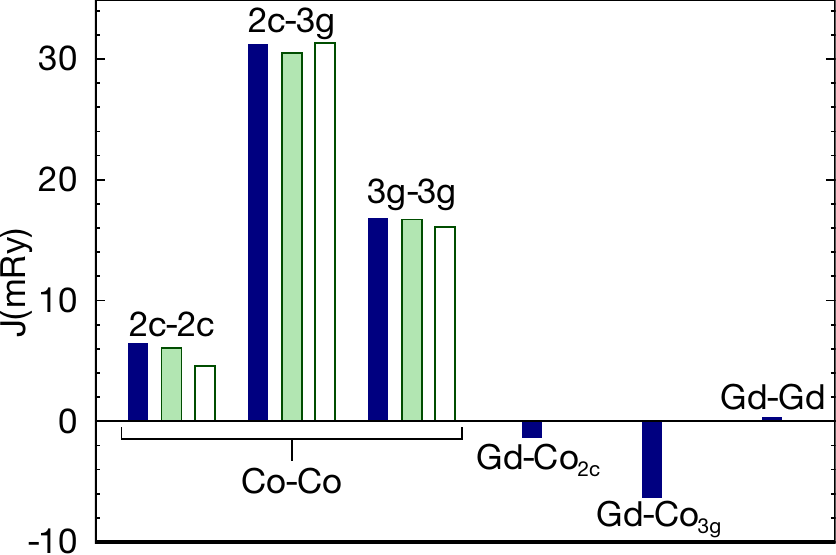}
\caption{
Calculated values of $J_{ij}$ in the high-temperature 
expansion of equation~\ref{eq.expansion}
for GdCo$_5$ (blue) and YCo$_5$ (green).
The empty bars are the values of $J_{ij}$ computed for YCo$_5$
using the lattice parameters of GdCo$_5$ (see text).
\label{fig.pristineJ}
}
\end{figure}

The relative strengths of the
TM-TM, RE-TM and RE-RE interactions can be quantified
by expanding the calculated potential energy 
$\langle \Omega \rangle_{0,T}$ in terms
of order parameters describing the thermally-averaged
local moment at the different sublattices ($m_i$; 
equation~\ref{eq.m_i}).
Close to $T_\mathrm{C}$ ($m_i\rightarrow0$) the expansion can be truncated at
second order, i.e.:
\begin{eqnarray}
\langle \Omega \rangle_{0,T}
&\approx&  \left(-\frac{1}{2} J_{2c-2c} m_{\mathrm{Co}_{2c}}^2 
- J_{2c-3g} m_{\mathrm{Co}_{2c}}m_{\mathrm{Co}_{3g}} \right.\nonumber \\
&&
\left. - \frac{1}{2} J_{3g-3g} m_{\mathrm{Co}_{3g}}^2 \right)
-\frac{1}{2} J_\mathrm{Gd-Gd} m_\mathrm{Gd}^2 
\nonumber \\
&& - J_{\mathrm{Gd-Co}_{2c}} m_\mathrm{Gd}m_{\mathrm{Co}_{2c}}
   - J_{\mathrm{Gd-Co}_{3g}} m_\mathrm{Gd}m_{\mathrm{Co}_{3g}} \nonumber \\
&&
\label{eq.expansion}
\end{eqnarray}
where we have decomposed the Co contribution into the two
inequivalent $2c$ and $3g$ sublattices (Fig.~\ref{fig.ballstick}),
and assumed collinear magnetization of the sublattices.
Only the terms in parentheses are required for YCo$_5$.
Differentiation of equation~\ref{eq.expansion} with respect
to $m_i$ yields expressions for the Weiss fields through
equation~\ref{eq.hdiff}, conveniently expressed in matrix form:
\begin{equation}
\begin{pmatrix}
h_{\mathrm{Co}_{2c}} \\
h_{\mathrm{Co}_{3g}} \\
h_\mathrm{Gd}
\end{pmatrix}
=
\begin{pmatrix}
\frac{J_{2c-2c}}{2} & \frac{J_{2c-3g}}{2} & \frac{J_{\mathrm{Gd-Co}_{2c}}}{2} \\
\frac{J_{2c-3g}}{3} & \frac{J_{3g-3g}}{3} & \frac{J_{\mathrm{Gd-Co}_{3g}}}{3} \\ 
J_{\mathrm{Gd-Co}_{2c}} & J_{\mathrm{Gd-Co}_{3g}}& J_\mathrm{Gd-Gd}        
\end{pmatrix}
\begin{pmatrix}
m_{\mathrm{Co}_{2c}} \\
m_{\mathrm{Co}_{3g}} \\
m_\mathrm{Gd}
\end{pmatrix}.
\label{eq.hmatrix}
\end{equation}
The denominators of 2 and 3 account for the 
multiplicities of the $2c$ and $3g$ positions.
We then obtain the $J_{ij}$ coefficients from a least-squares fit of
the calculated $\{h_i\}$ values from a training set of 
$\{m_i\}$ (equivalently, $\{ \lambda_i\}$),
and plot them in Fig.~\ref{fig.pristineJ}.
It is essential to stress that the $J_{ij}$ values
are not simply describing 
pairwise interactions, but rather should be thought of as
coefficients in the rather general expansion of $\langle \Omega \rangle_{0,T}$ 
in equation~\ref{eq.expansion}.
This point is discussed further in Ref.~\citenum{MendiveTapia2017}.

Initially focusing on GdCo$_5$ (blue bars in Fig.~\ref{fig.pristineJ}),
we first note the negative values of 
$J_{\mathrm{Gd-Co}_{2c}}$ and $J_{\mathrm{Gd-Co}_{3g}}$, as expected
for antiferromagnetic alignment.
The RE-RE interaction quantified by  $J_\mathrm{Gd-Gd}$ is ferromagnetic
but negligibly small, i.e.\ the RE ordering is driven
by RE-TM interactions.
Interestingly, $J_{\mathrm{Gd-Co}_{3g}}$ is 4.5 times 
larger than $J_{\mathrm{Gd-Co}_{2c}}$, showing that
the dominant RE-TM interaction is not between in-plane
nearest neighbors, but rather between the RE and
the adjacent pure Co planes.
It follows that substituting Co at the $3g$ positions
should have a greater effect on the RE than at the $2c$ positions,
a hypothesis that we test in Sec.~\ref{sec.dopeGd}.

Turning to the TM-TM interaction in GdCo$_5$, again we find
the largest $J_{ij}$ to correspond to interplanar interactions,
i.e.\ $J_{2c-3g}$.
The in-plane interactions $J_{2c-2c}$, $J_{3g-3g}$ are also 
ferromagnetic but smaller by $ J_{2c-3g}$ by factors of 5 and 2,
respectively.
Comparing these $J_{ij}$ values with those found for
YCo$_5$ (green filled bars in Fig.~\ref{fig.pristineJ})
we find the same ordering of values and similar magnitudes,
but the dominant $J_{2c-3g}$ coefficient of GdCo$_5$ 
is larger by 2.4\%.

Given that the values of $J_{ij}$ determine $T_\mathrm{C}$ (discussed
in the following section), we investigated the
origin of the difference in $J_{2c-3g}$ by performing a 
calculation on YCo$_5$ using the lattice parameters of GdCo$_5$.
This procedure amounts to increasing the $a$ parameter by 0.5\%
and reducing the $c$ parameter by 0.4\%.\cite{Andreev1991,AndreevHMM}
The resulting $J_{ij}$ values are shown as the empty green
bars in Fig.~\ref{fig.pristineJ}.
We see that the respective increase and decrease in $a$ and $c$ 
coincide with weakened in-plane interactions ($J_{2c-2c}$, $J_{3g-3g}$).
However,  the interplanar interaction is strengthened by 2.9\%, leading us
to attribute the difference in $J_{2c-3g}$ between GdCo$_5$ and YCo$_5$
to be structural in origin.
We surmise that the RE can indirectly modify the TM-TM interaction
through chemical pressure.

\subsection{Calculation of $T_\mathrm{C}$ from $J_{ij}$}
\label{sec.TcJ}
Equation~\ref{eq.hmatrix} can be used to calculate $T_\mathrm{C}$
by replacing $m_i = L(\lambda_i) = L(\beta h_i)$ and
using the $m_i \rightarrow 0$ limit,
$L(x) \rightarrow \frac{x}{3}$.
Equation~\ref{eq.hmatrix} then reduces to an eigenvalue
problem, with the smallest $\beta$ corresponding to $T_\mathrm{C}$.
This approach allows the analysis of the difference in $T_\mathrm{C}$
between GdCo$_5$ and YCo$_5$.
For instance, taking the $J_{ij}$ values obtained for YCo$_5$
and then replacing $J_{2c-3g}$ with the larger value obtained for
GdCo$_5$ increases the calculated $T_\mathrm{C}$ from 885~K to 900~K.
Further replacing $J_{2c-2c}$ and $J_{3g-3g}$ gives a further
increase in $T_\mathrm{C}$ to 906~K.

It follows that the remaining 60\% of the increase in $T_\mathrm{C}$ observed for
GdCo$_5$ (34~K, to 940~K) must be attributed to
the RE-TM and/or RE-RE interaction.
We find that the small value of $J_\mathrm{Gd-Gd}$ means 
that the RE-RE interaction accounts for less than 1~K of the difference,
so it is the RE-TM interaction, especially the interplanar
interaction characterized by $J_{\mathrm{Gd-Co}_{3g}}$, which
is responsible.
Therefore according to the calculations, although the RE-TM
interaction does not affect the Co sublattice magnetization
below 400~K (Fig.~\ref{fig.GdCo5resolved}), the interaction
is essential to understanding the higher $T_\mathrm{C}$ of GdCo$_5$.

\subsection{Weiss field on Gd}
\label{sec.WeissGd}

\begin{figure}
\includegraphics{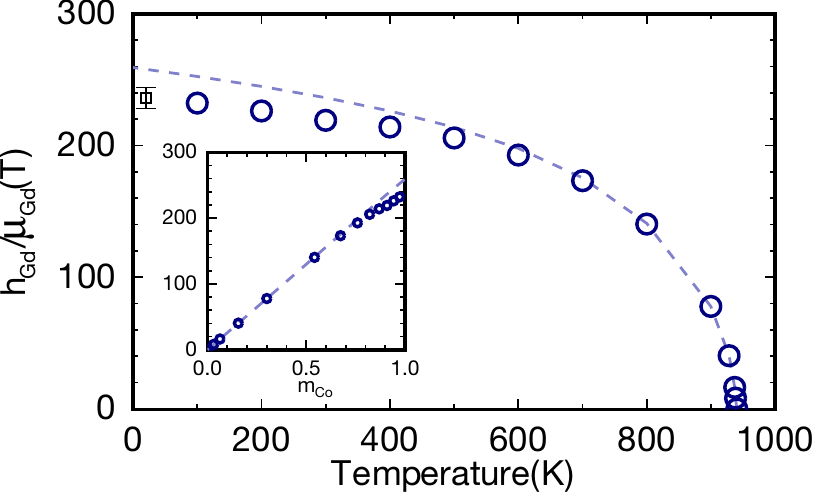}
\caption{
The molecular field on Gd in GdCo$_5$.
The open circles show the calculated Weiss fields
divided by the local moment magnitudes as a function
of temperature.
The inset contains the same data as a function
of the average Co order parameter.
The dashed line shows the expected Weiss field
based on the $J$ expansion of equation~\ref{eq.expansion}
and parameters shown in Fig.~\ref{fig.pristineJ}.
The open square with error bars in the main panel 
denotes the molecular field
measured by inelastic neutron scattering experiments 
at 20~K as reported in Ref.~\citenum{Loewenhaupt1994}.
\label{fig.hGd}
}
\end{figure}

In Fig.~\ref{fig.hGd} we plot the temperature evolution
of $h_\mathrm{Gd}$, the calculated Weiss field on Gd
in GdCo$_5$.
Since $h_i$ has units of energy (equation~\ref{eq.omega0})
we convert to a field in tesla by dividing by the calculated
local moment magnitude $\mu_\mathrm{Gd}$, which varies from
7.30~to~7.05$\mu_B$ from $T=0$~K to $T_\mathrm{C}$.
The inset plots the same data against the averaged
Co order parameter, $m_\mathrm{Co} = (2m_{\mathrm{Co}_{2c}} 
+3m_{\mathrm{Co}_{3c}})/5$.

The dashed line in Fig.~\ref{fig.hGd} 
shows the expected behavior of
$h_\mathrm{Gd}$ according to equation~\ref{eq.hmatrix}.
By construction this fit is accurate close to $T_\mathrm{C}$,
but at temperatures below 600~K deviations are observed,
such that $h_\mathrm{Gd}$ is no longer linear in $m_\mathrm{Co}$
(inset).
To accurately reproduce the calculated Weiss field at the
RE site at these temperatures it is necessary to include
higher-order terms\cite{MendiveTapia2017} in the 
expansion of equation~\ref{eq.expansion},
preventing a straightforward mapping to a Heisenberg-like
Hamiltonian.

Although the Weiss fields were introduced as parameters
as a means of modeling the local moment statistics, it is
reasonable to ask how they compare to the exchange
field at the RE site which can be measured via inelastic neutron
scattering (INS).\cite{Loewenhaupt1994}
Therefore in Fig.~\ref{fig.hGd} we also plot the value
of 236$\pm$8~T at 20~K which was measured in the INS experiments
of Ref.~\citenum{Loewenhaupt1994}.
The excellent agreement with the calculated values of 
$h_\mathrm{Gd} / \mu_\mathrm{Gd}$ is perhaps fortuitous
and certainly sensitive to the spherical approximation to
the potential,\cite{Kuzmin2004} but nonetheless
gives us confidence in the validity of the local
moment description of the RE magnetism.

\subsection{Substitutional doping of transition metals I:
TM sites}
\label{sec.firstdope}

We now go beyond the pristine RECo$_5$
compounds and consider substitutional doping of the transition
metals.
We have investigated both experimentally and computationally 
the replacement of Co with its
neighboring elements Fe and Ni, considering the compounds
RECo$_{4.5}$Ni$_{0.5}$, RECo$_4$Ni and  
RECo$_{4.5}$Fe$_{0.5}$.
These low dopant concentrations were chosen to avoid
complications arising from structural modification
through doping\cite{AndreevHMM} and the low solubility of 
Fe.\cite{Chuang1982,Taylor1975} 
Even so, due to this low solubility we were unable to synthesize
a single-phase sample of GdCo$_{4.5}$Fe$_{0.5}$.

Previous experimental studies\cite{Deportes1976,
Laforest1973,Rothwarf1973} attempted to determine whether
the dopants preferentially occupy $2c$
or $3g$ sites (Fig.~\ref{fig.ballstick}) or are distributed
equally among the TM sublattices.
The neutron diffraction experiments of Ref.~\citenum{Deportes1976}
on Ni-doped YCo$_5$ found a preference for Ni substitution
at $2c$ sites (with $2c$/$3g$ occupancies of 0.16/0.06 for
YCo$_{4.5}$Ni$_{0.5}$ and 0.29/0.14 YCo$_4$Ni).
For Fe-doped YCo$_5$ we are unaware of similar neutron measurements,
but the study of the related compound ThCo$_5$ in Ref.~\citenum{Laforest1973} 
found a preference for Fe-substitution at 3$g$ sites
($2c$/$3g$ occupancies of 0.2/0.5 for YCo$_3$Fe$_2$).
On the other hand Ref.~\citenum{Rothwarf1973} argued that
the evolution of lattice parameters of YCo$_5$ 
as a function of
Fe-doping was consistent with preferential substitution
at $2c$ sites.

\begin{figure}
\includegraphics{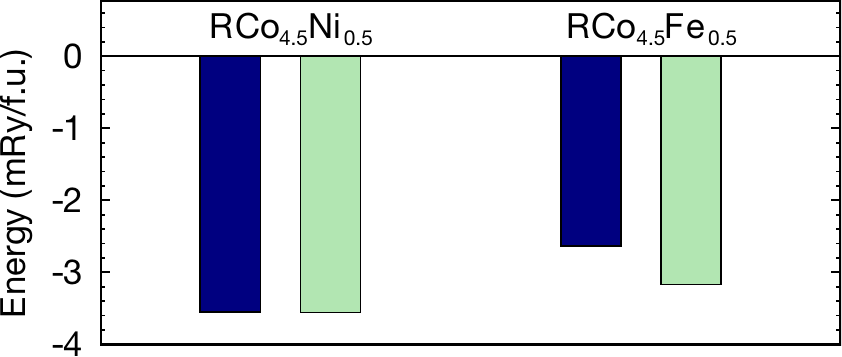}
\caption{
Calculated energetics of doping of GdCo$_5$ (blue)
or YCo$_5$ (green) by substituting at a Co$_{2c}$ site.
The $y$-axis zero corresponds to the energy per formula
unit when the dopant is substituted at a Co$_{3g}$ site,
i.e.\ negative bars imply the dopant is more stable
sitting at a Co$_{2c}$ site.
\label{fig.dopeenergy}
}
\end{figure}

We have calculated the ground-state (zero temperature) energies
of RECo$_{4.5}$T$_{0.5}$,
T = Ni or Fe, where the dopants were substituted either
on the $2c$ or $3g$ sites.
The energy differences per formula unit between the two cases 
for RE~=~Gd and Y
are shown in Fig.~\ref{fig.dopeenergy}.
The negative values displayed in Fig.~\ref{fig.dopeenergy} 
imply that, according to our CPA calculations,
$2c$-substitution is more stable for both Ni and Fe-doping
of both GdCo$_5$ and YCo$_5$ (blue and green bars)
Interestingly, there is a notable difference in the energetics 
of Fe-doping between GdCo$_5$ and YCo$_5$.
As discussed in Sec.~\ref{sec.dopeGd} this difference 
is due to a magnetic energy penalty in placing Fe 
at $2c$ sites when Gd is present.

Although the CPA calculations support $2c$-ordering,
the different conclusions drawn based on
experiments\cite{Laforest1973,Rothwarf1973} may 
indicate a dependence on sample preparation
routes.
Therefore in order to keep our study general, in the following
we present calculations for both $2c$ and $3g$ preferential doping.
We view these calculations as limiting cases, with the 
experimentally-realized situation lying somewhere 
in between.

\subsection{Substitutional doping of transition metals II:
Magnetizations}
\label{sec.MvH}
\begin{figure}
\includegraphics{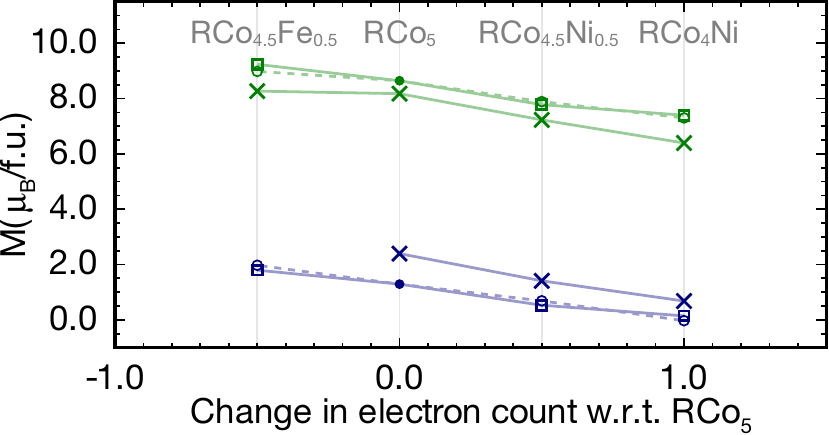}
\caption{
Low temperature magnetizations of doped RECo$_5$ compounds.
The crosses show the magnetization of powdered samples
in a field of 7~T at 5~K for Fe or Ni-doped YCo$_5$ (green)
and GdCo$_5$ (blue).
The circles and squares are the magnetizations calculated
where the dopants have
been substituted either at Co$_{2c}$
or Co$_{3g}$  sites respectively.
\label{fig.dopemag}
}
\end{figure}

In Fig.~\ref{fig.dopemag} we present the saturation
magnetizations measured and calculated for the doped RECo$_5$
compounds.
As we might expect, the behavior with doping of GdCo$_5$
and YCo$_5$ is very similar.
The general trend is of an increase in magnetization
with Fe-doping and a decrease with Ni-doping.
This behavior is consistent with a rigid-band picture,
noting that in YCo$_5$ the $d$-band is essentially 
full in the majority-spin channel and partially occupied
in the minority channel;\cite{Koudela2008} therefore
increasing the electron count (through Ni-doping) further
populates the minority-spin channel and decreases the overall
moment, and vice versa for Fe-doping.
The calculated magnetizations for the dopants occupying
$2c$ or $3g$ sites (circles and squares in Fig.~\ref{fig.dopemag})
are very similar.
The supercell calculations of Ref.~\citenum{Larson2004} found
the same behavior, again consistent with the rigid band
model.

\begin{figure}
\includegraphics{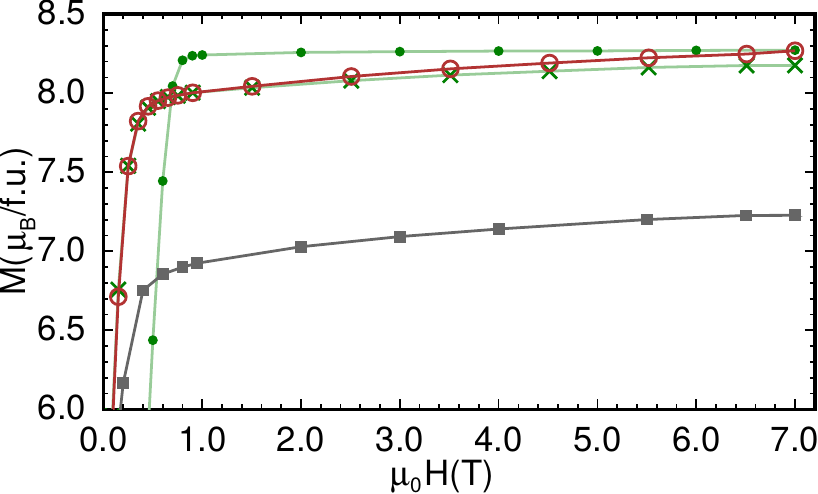}
\caption{
Magnetization vs.\ applied field measured at 5~K for single crystal YCo$_5$
(green line, filled circles), polycrystalline (powdered) YCo$_5$ (green line, crosses), 
YCo$_{4.5}$Ni$_{0.5}$ (gray line, squares)  and YCo$_{4.5}$Fe$_{0.5}$ 
(brown line, empty circles).
\label{fig.MvH}
}
\end{figure}
We now compare the magnetic moments for the polycrystalline (powdered) samples of 
the pristine compounds (YCo$_5$ and GdCo$_5$) presented in Fig.~\ref{fig.dopemag} with 
the values obtained for the magnetic moments of the single crystals given in Table~\ref{tab.zeroKM}.
For example, we note that the moment value for the polycrystalline YCo$_5$ is 
0.23$\mu_B$/f.u.\ lower than the value obtained for the YCo$_5$ single crystal. 
In order to explain this small difference, we focus our attention on the isothermal 
magnetization plots shown in Fig.~\ref{fig.MvH} obtained at $T = 5$~K for all the 
polycrystalline (Fe, Ni)-doped YCo$_5$ samples, where for comparison we also 
plot the magnetization of the YCo$_5$ single crystal (green line).
During the measurement process, the magnetic field was reduced from 7 to 0~T and 
the magnetization data were recorded at several field values.
It is apparent that none of the $M$v$H$ curves for the polycrystalline materials 
saturate, even at a field of 7~T.
In contrast, the $M$v$H$ curve for the single crystal saturates above $\mu_0H = 1$~T. 
This demonstrates that it is easier to saturate the magnetization of a single crystal 
(when $H$ is applied along the easy axis of magnetization).
For a polycrystalline sample of doped or pure YCo$_5$ made up of a collection of 
randomly aligned grains (with randomly aligned easy axes of magnetization), the 
magnetization at any field below the anisotropy field will provide a lower bound 
on the saturation magnetization.
For GdCo5, the situation is even more complex due to its ferrimagnetic ordering, 
which can lead to non-collinear Gd and Co spins when the applied field is not 
parallel to the easy axis.\cite{Radwanski1986}
We have also observed that using solid rather than powder polycrystalline samples of 
YCo$_5$ leads to even lower values for the magnetic moment at the same $H$ and $T$ 
(data not shown here).
Nevertheless, using powder samples one can obtain data that 
can be used to identify trends, e.g. the variation in the saturation magnetization 
with doping within a sample series, and the saturation moments 
obtained lie within a few percent of the single crystal values.

\subsection{Substitutional doping of transition metals III:
$T_\mathrm{C}$}

\begin{figure}
\includegraphics{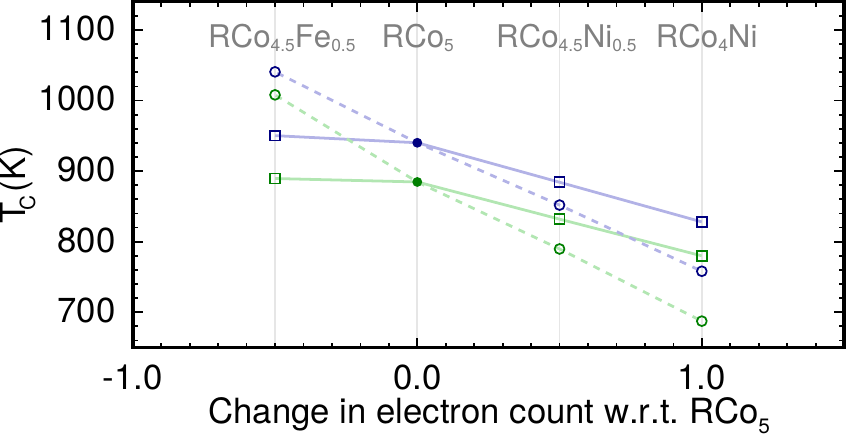}
\caption{
Curie temperatures calculated for
YCo$_5$ (green) and GdCo$_5$ (blue) for different
doping concentrations, where the dopants have
been substituted either at Co$_{2c}$ (circles, dashed lines)
or Co$_{3g}$ (squares, solid lines) sites.
\label{fig.dopeTc}
}
\end{figure}

\begin{figure*}
\includegraphics{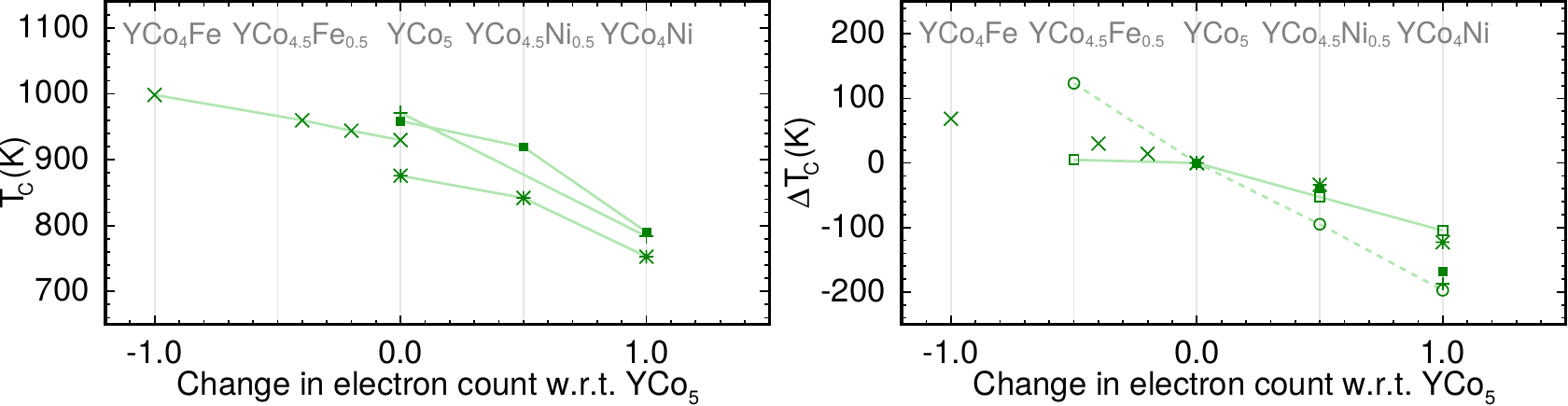}
\caption{
Curie temperatures of doped YCo$_5$, shown on an absolute
scale (left panel) or relative to the $T_\mathrm{C}$ measured/calculated
for pristine YCo$_5$ (right panel).
The experimental $T_\mathrm{C}$ values were previously reported
in Refs.~\citenum{Paoluzi1994}
(diagonal crosses),~\citenum{Ermolenko1977} (filled squares),
~\citenum{Chuang1982} (upright crosses) and~\citenum{Buschow1976} (asterisks).
The right panel additionally shows the calculated $T_\mathrm{C}$ values for
doped YCo$_5$ (cf.\ Fig.~\ref{fig.dopeTc}) with the dopants
at Co$_{2c}$ (circles, dashed lines)
or Co$_{3g}$ (empty squares, solid lines) sites.
\label{fig.dopeTcexpt}
}
\end{figure*}

In Fig.~\ref{fig.dopeTc} we present the calculated Curie temperatures
for the doped compounds.
The variations in $T_\mathrm{C}$ with doping are found to be very similar for
RE=Gd and Y, displaying the same $\sim$60~K offset as observed for
the pristine case and discussed in Sec.~\ref{sec.TcJ}.
However, unlike the magnetization plotted in Fig.~\ref{fig.dopemag},
the $T_\mathrm{C}$ values show a pronounced dependence on whether the dopants
are substituted at the $2c$ or $3g$ sites.
The largest variations in $T_\mathrm{C}$ occur when the dopants occupy
the $2c$ sites, e.g.\ increasing by 124~K for YCo$_{4.5}$Fe$_{0.5}$
and decreasing by 95~K for YCo$_{4.5}$Ni$_{0.5}$.
However,  doping with Fe on the $3g$ sites only raises $T_\mathrm{C}$ by 5~K
for YCo$_{4.5}$Fe$_{0.5}$.

Further insight into the behavior of $T_\mathrm{C}$ can be obtained
by extending the analysis of Sec.~\ref{sec.orderparam}.
The appropriate modification of equation~\ref{eq.hmatrix} is
\begin{widetext}
\begin{equation}
\begin{pmatrix}
h_{\mathrm{Co}_{2c}} \\
h_{\mathrm{Co}_{3g}} \\
h_\mathrm{T} \\
h_\mathrm{Gd}
\end{pmatrix}
=
\begin{pmatrix}
\frac{c_{2c}J_{2c-2c}}{2} & \frac{c_{3g}J_{2c-3g}}{2} & \frac{c_{T}J_{2c-T}}{2}&\frac{J_{\mathrm{Gd-Co}_{2c}}}{2} \\
\frac{c_{2c}J_{2c-3g}}{3} & \frac{c_{3g}J_{3g-3g}}{3} & \frac{c_{T}J_{3g-T}}{3}&\frac{J_{\mathrm{Gd-Co}_{3g}}}{3}  \\ 
\frac{c_{2c}J_{2c-T}}{n}  & \frac{c_{3g}J_{3g-T}}{n} & \frac{c_{T}J_{T-T}}{n}&\frac{J_\mathrm{Gd-T}}{n}  \\ 
c_{2c}J_{\mathrm{Gd-Co}_{2c}} &c_{3g} J_{\mathrm{Gd-Co}_{3g}}&c_{T}J_{\mathrm{Gd-T}}& J_\mathrm{Gd-Gd}        
\end{pmatrix}
\begin{pmatrix}
m_{\mathrm{Co}_{2c}} \\
m_{\mathrm{Co}_{3g}} \\
m_{\mathrm{T}} \\
m_\mathrm{Gd}
\end{pmatrix}
\label{eq.hmatrixdoped}
\end{equation}
\end{widetext}
where $n$ is the multiplicity
of the dopant sites (2 or 3 for $2c$ or $3g$ doping, respectively).
Removing all terms involving Gd gives the expression for YCo$_5$.
For the compound RECo$_{5-x}$T$_x$, the dopant concentration $c_T$ is
given by $x/n$, while the Co concentrations
$(c_{2c},c_{3g})$  equal $(1 - c_T,1)$ for $2c$-doping and vice versa
for $3g$-doping.

We proceed as in Sec.~\ref{sec.TcJ} to obtain the $J_{ij}$ values and $T_\mathrm{C}$.
Postponing a discussion of GdCo$_5$ to the next section, this analysis
for YCo$_5$ reveals two key points.
First, for Ni-doping, $J_{2c-2c}$, $J_{2c-3g}$ and $J_{3g-3g}$
only undergo small changes from the pristine case, while the $J$
parameters coupling to Ni are negligible.
Therefore the observed reduction in $T_\mathrm{C}$ with Ni-doping is essentially
a dilution effect.
We recall from  Fig.~\ref{fig.pristineJ} that the interlayer coupling
dominates the magnetic properties.
Doping on the $2c$ site therefore has a larger effect on $T_\mathrm{C}$
simply due to the lower multiplicity of this site; taking YCo$_4$Ni
as an example, $2c$-doping reduces the cobalt content in a layer
by 50\% compared to only 33\% with $3g$-doping.
This difference alone can account for a 20~K reduction in $T_\mathrm{C}$ 
moving from $3g$ to $2c$-doping.

The second point applies to the Fe-doped compound YCo$_{4.5}$Fe$_{0.5}$.
When solving the eigenvalue problem of equation~\ref{eq.hmatrixdoped},
the eigenvectors give the relative ordering strengths of the different
sublattices.
For the cases of $2c$ and $3g$-doping respectively, the normalized
$(h_{\mathrm{Co}_{2c}}, h_{\mathrm{Co}_{3g}},h_\mathrm{Fe})$ eigenvectors
are (0.49,0.44,0.75) and (0.61,0.55,0.58).
That is, for $2c$-doping the magnetic ordering close to $T_\mathrm{C}$ is dominated
by the Fe sublattice, thanks to a large value of $J_\mathrm{Fe-Fe}$ (29~mRy).
As we explore in the next section, the presence of Fe at the $2c$ sites
also modifies the exchange field at the RE site.

In Fig.~\ref{fig.dopeTcexpt} we compare our calculated $T_\mathrm{C}$ for YCo$_{5-x}$T$_x$
with previously-published experimental data.\cite{Paoluzi1994,Ermolenko1977,Chuang1982,Buschow1976}
The experiments also find an increase or decrease in $T_\mathrm{C}$ for Fe or Ni-doping, respectively.
As already noted, the calculated $T_\mathrm{C}$ for YCo$_5$ is lower than that measured
experimentally, and the left panel of Fig.~\ref{fig.dopeTcexpt} also illustrates the scatter
in reported experimental values.
Therefore in the right panel of Fig.~\ref{fig.dopeTcexpt} we plot the same data as a difference
relative to the $T_\mathrm{C}$ measured for YCo$_5$, and include our calculated
data for $2c$ or $3g$-doping.
With the exception of YCo$_{4.5}$Ni$_{0.5}$ the experimental data points
fall in between the $2c$/$3g$ limiting cases.
We tentatively note that the values of $T_\mathrm{C}$ 
of Fe-doped YCo$_5$ measured in Ref.~\citenum{Paoluzi1994} do not
show the large increase predicted for preferential $2c$ substitution,
which would support the conclusion based on ThCo$_5$ that
$3g$ substitution is preferable.\cite{Laforest1973}
However,  given the uncertainties in measuring and calculating $T_\mathrm{C}$
we acknowledge that such an indirect assignment can only be speculative.

\subsection{Substitutional doping of transition metals IV:
Modification of the RE-TM interaction through doping}
\label{sec.dopeGd}
\begin{figure*}
\includegraphics{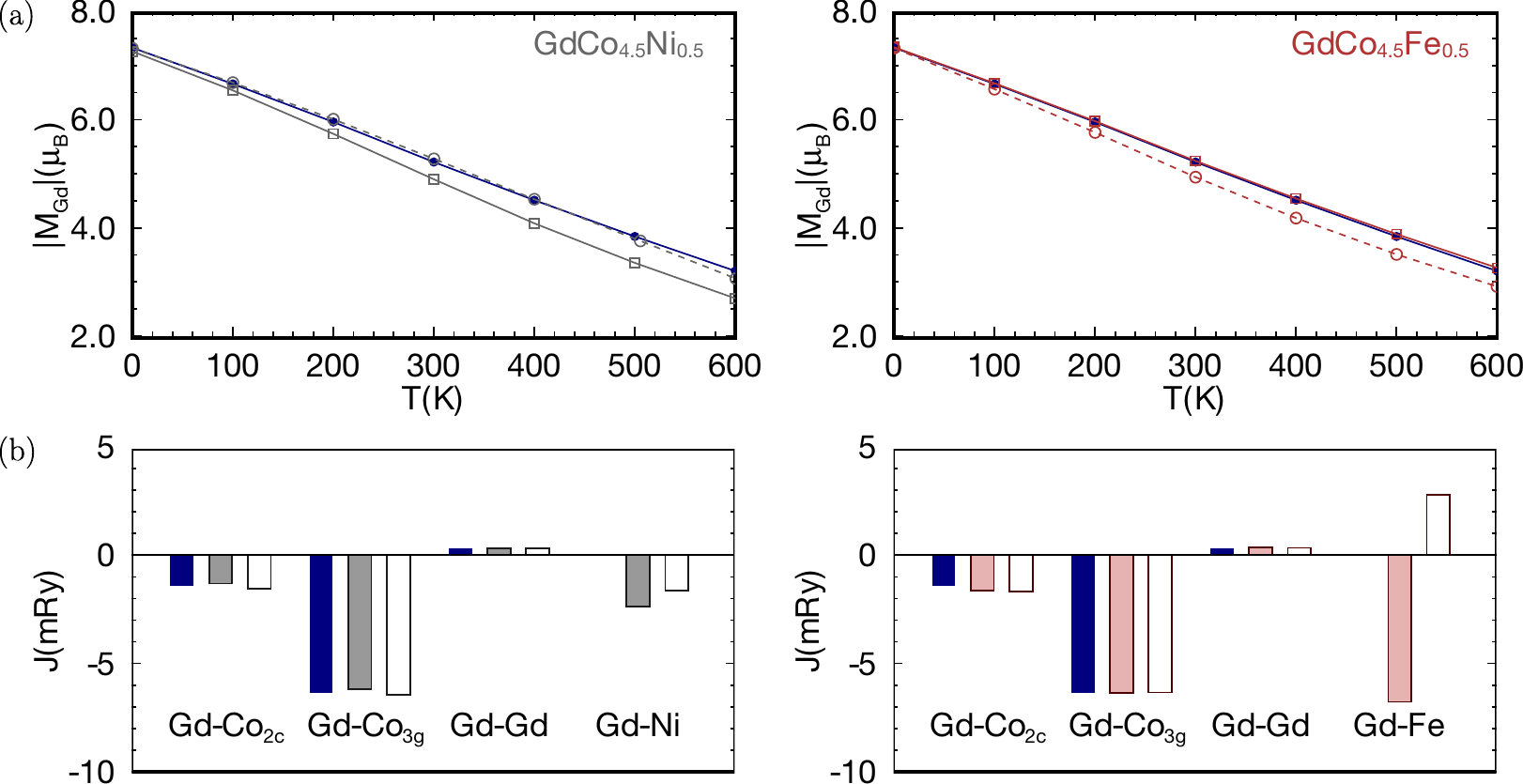}
\caption{
(a) Absolute magnetization of the Gd sublattice and
(b) calculated $J$ parameters
for Ni or Fe-doped GdCo$_{4.5}$T$_{0.5}$.
Doping on the Co$_{2c}$ or Co$_{3g}$ site is 
denoted in (a) by empty circles and squares, respectively,
and by empty and light-filled bars in (b).
Dark blue filled circles/bars correspond to pristine GdCo$_5$.
\label{fig.dopeJ}
}
\end{figure*}

Aside from modification of the magnetization and $T_\mathrm{C}$, it is
important to establish the effect that substitutional doping
has on the RE.
For instance, since it is the Sm itself which provides
the large anisotropy in SmCo$_5$,\cite{Ermolenko1976} control
of the RE is equivalent to controlling the anisotropy.
For the current case, it is important to establish whether
the difficulty in synthesizing GdCo$_{4.5}$Fe$_{0.5}$ has a
magnetic origin.
Therefore we use our calculations to investigate the 
RE-TM interaction in 
GdCo$_{4.5}$T$_{0.5}$ for T = Ni, Fe.
In Fig.~\ref{fig.dopeJ}(a) we show the temperature
evolution of the Gd magnetization 
(cf.\ Fig.~\ref{fig.GdCo5resolved} for pristine GdCo$_5$)
for preferential $2c$ or $3g$-doping.
In Fig.~\ref{fig.dopeJ}(b) we plot the calculated $J_{ij}$
parameters of equation~\ref{eq.hmatrixdoped} which quantify
the RE-TM interaction.

Focusing on Ni-doping first (left panels of Fig.~\ref{fig.dopeJ})
we find that doping on the $2c$ site has a negligible effect 
on the Gd magnetization.
Indeed, we find the value of $J_{\mathrm{Gd-Ni}}$ to be 
close to $J_{\mathrm{Gd-Co}_{2c}}$, despite the weaker
magnetism of Ni.
However, doping with Ni on the $3g$-site reduces 
the exchange field at the RE site and causes a faster
reduction in the Gd magnetization with temperature.
Although the value of $J_{\mathrm{Gd-Ni}}$ 
calculated for $3g$-doping is larger than that calculated 
for $2c$-doping, it is 
smaller than $J_{\mathrm{Gd-Co}_{3g}}$ by almost
50\%.
Given that it is $J_{\mathrm{Gd-Co}_{3g}}$ which drives
the RE ordering (Sec.~\ref{sec.orderparam}),
this reduction has a noticeable effect on the RE
magnetization.

Given that Ni is magnetically weaker than Co, 
it is not too surprising that we observe a weaker RE-TM 
interaction.
Conversely, given that both $T_\mathrm{C}$ and the zero temperature
magnetization increase with Fe-doping, it is tempting to
assume that Fe-doping might strengthen the RE-TM interaction,
especially when substituted at $3g$ sites.
However, our calculations (right panel of Fig.~\ref{fig.dopeJ})
do not support this view.
Doping at the $3g$ site does give a slightly slower
decay of the Gd magnetization due to an enhanced 
value of $J_{\mathrm{Gd-Fe}}$.
However, this value is only 6\% larger than 
$J_{\mathrm{Gd-Co}_{3g}}$ [filled red bars 
in Fig.~\ref{fig.dopeJ}(b)], so in GdCo$_{4.5}$Fe$_{0.5}$
the effect is minimal.

Surprisingly, our calculations further find
that Fe-doping at the $2c$-site actually 
weakens the RE-TM interaction and causes a faster
temperature decay of the Gd magnetization compared
to the pristine case
[right panel of Fig.~\ref{fig.dopeJ}(a)].
This unexpected result can be traced to a \emph{positive}
value of $J_{\mathrm{Gd-Fe}}$, i.e.\ a ferromagnetic
interaction between the RE and the Fe atoms
located at the $2c$ sites.
This finding is robust against the choice of spherical approximation
to the potential (using the muffin-tin approximation).
We note that such a ferromagnetic interaction cannot
be accounted for in the standard model of RE-TM
interactions based on the hybridization of minority
TM-$3d$ with majority RE-$5d$ electrons.\cite{Kumar1988}
The fact that this behavior is only calculated for 
$2c$-doping indicates the existence of a secondary effect
when the Fe dopants are placed at nearest neighbor positions
to the RE.
Such competing magnetic interactions will have a detrimental
effect on the solubility of Fe.
It is interesting to note that codoping GdCo$_5$ with B
stabilizes compounds with higher Fe content, given that B occupies precisely
these $2c$ sites.\cite{Drzazga1989}

\section{Summary and conclusions}
\label{sec.conclusions}

We have studied the RECo$_{5-x}$T$_x$ family
of compounds where RE = Y and Gd and T = Ni and Fe.
Our purpose was to probe
the TM-TM and RE-TM interactions which govern
rare-earth/transition-metal permanent magnets,
taking advantage of the relatively simple 
RECo$_5$ crystal structure and lack of
crystal-field interactions.
We have combined state-of-the-art computational and
experimental methods: first-principles
calculations based on self-interaction corrected DFT
and the disordered local moment picture
to calculate magnetic properties for $0<T<T_\mathrm{C}$, 
and single-crystal
growth with the optical floating zone technique
to obtain high-quality samples.

Beginning with the pristine YCo$_5$ and GdCo$_5$
compounds, we obtained a theoretical interpretation
of the experimentally-measured magnetization vs temperature curves.
In particular, the calculations explain the
opposite temperature dependences of the two compounds
and the ordering of $T_\mathrm{C}$.
The increase in GdCo$_5$ magnetization with temperature
was shown to arise from a faster decay of the Gd magnetization
compared to Co, while
the higher $T_\mathrm{C}$ of GdCo$_5$ was attributed to both
a modification of the lattice parameters due to the presence
of Gd, and also the favorable magnetic coupling between Gd and the
Co sublattices.
Expanding the potential energy in terms of order parameters
showed the dominant magnetic interaction to occur between
the planes of the hexagonal CaCu$_5$ structure.
Comparison of the calculated Weiss fields with the exchange
field at the RE site reported from INS
measurements\cite{Loewenhaupt1994} found good agreement, 
supporting the application of the DLM picture to this system.

For the doped systems, both experiments and calculations
showed an increase or decrease in magnetization with
Fe or Ni-substitution, respectively.
The calculations found that this change in magnetization
did not depend on whether the dopants were placed at the
$2c$ or $3g$ crystallographic sites.
The calculated values of $T_\mathrm{C}$ also showed the same increase/decrease
for Fe/Ni-doping, in agreement with previously-published data
for YCo$_{5-x}$T$_x$.\cite{Paoluzi1994,Ermolenko1977,Chuang1982,Buschow1976}
However,  here a dependence on the doping site was observed,
with larger changes in $T_\mathrm{C}$ calculated for $2c$-doping.
For Ni-doping this dependence was explained as a dilution effect,
while for Fe-doping the higher $T_\mathrm{C}$ for the $2c$ case was
found to arise from a strong Fe-Fe ferromagnetic interaction.

Examining the RE-TM interaction for the doped GdCo$_{5-x}$T$_x$
compounds, substituting Ni at the $3g$ site was found to
induce a faster reduction in the Gd magnetization with temperature,
as compared to the pristine compound or $2c$-doping.
However,  substituting Fe also showed this faster reduction
in magnetization, this time for $2c$-doping.
The order parameter expansion of the potential
energy surface traced the origin of this effect to
a ferromagnetic coupling between Gd and Fe at the
$2c$ sites.

Aside from these specific findings described above, the current study 
has laid the necessary groundwork
for the further investigation of the full RECo$_5$ family
(e.g.\ SmCo$_5$), where the RE-CF interactions play a key role.
In particular we have established the viability of
the experimental and computational protocols needed to synthesize,
characterize and model the RETM$_5$ permanent magnets.
However,  our study has also identified a new avenue of study
for GdCo$_{5-x}$Fe$_x$ regarding the Gd-Fe($2c$) interaction.
We have raised the possibility that the experimentally-known\cite{Drzazga1989}
necessity of codoping GdCo$_{5-x}$Fe$_x$ with B is related
to the calculated competition
between ferro and antiferromagnetic RE-TM interactions.
For Ni-substitution, although in the current study we have
focused on low doping concentrations, by extrapolating the GdCo$_5$
data in Fig~\ref{fig.dopemag} to higher Ni-doping we can expect 
a switch from TM to Gd-dominated
magnetization at zero temperature, which should yield a compensation point.
There is also a question of whether the TM-magnetization collapses
at a critical concentration of Ni or whether it continuously decreases
to zero.\cite{Ishikawa2003}

As a final note, we point out that the current study has
focused on magnetization along
a single direction and not addressed anisotropic 
quantities.
Aside from the study of pristine YCo$_5$ presented in Ref.~\citenum{Matsumoto2014},
there is further work to be done regarding the doped compounds.
More fundamentally there is the question of the anomalous temperature
dependence of the magnetocrystalline anisotropy in GdCo$_5$, particularly
regarding the role of anisotropic exchange.\cite{Gerard1992,Radwanski1992,
Alameda1981} 
Through the combination of our fully-relativistic calculations 
with high-quality single crystals, we are well-equipped to address such 
questions in future work.

\begin{acknowledgments}
The present work forms part of the PRETAMAG project, 
funded by the UK Engineering and Physical Sciences Research Council, 
Grant no. EP/M028941/1.
Work at Daresbury Laboratory was supported by an EPSRC service level 
agreement with the Scientific Computing Department of STFC.
We acknowledge useful discussions with M.\ Matsumoto, 
Prof.\ G.\ Rowlands,
M.\ Laver, M.\ Lueders, Z.\ Szotek and A.\ Walton.
We thank Dr.\ A.\ Vasylenko for his assistance in
translating Ref.~\citenum{Ermolenko1977}, 
Dr.\ M.\ Ciomaga Hatnean for assistance with single
crystal growth and
D.\ A.\ Duncan and O.\ J.\ Parish for preparing initial
samples of doped RECo$_5$ compounds.
\end{acknowledgments}

\appendix
\section{Structural characterization}
\label{app.struc}

\begin{figure}
\includegraphics{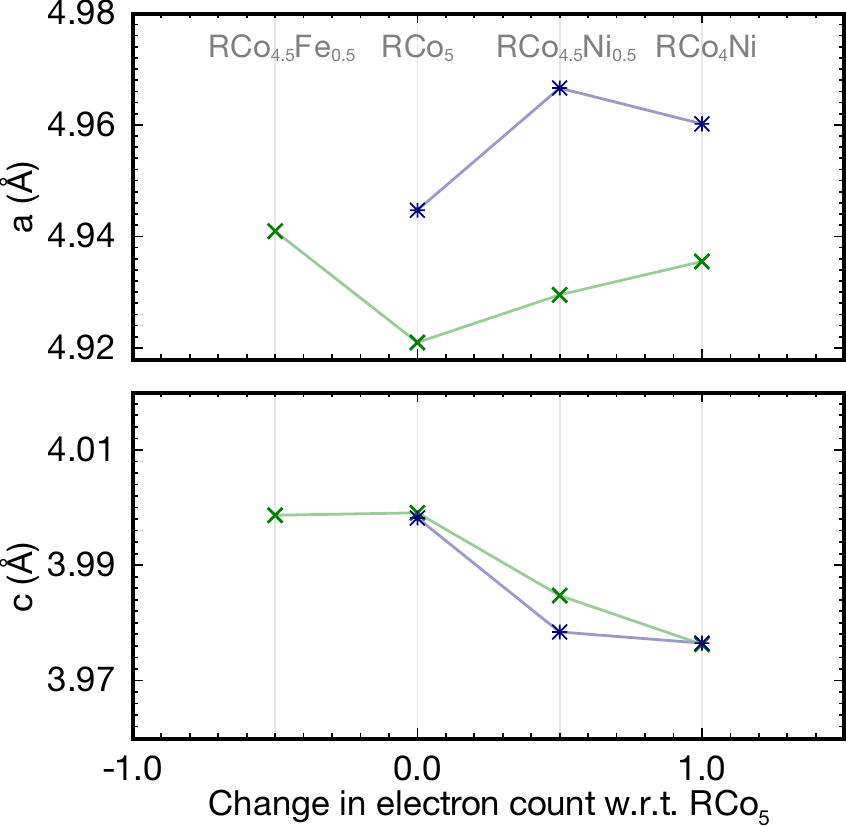}
\caption{Lattice parameters (in \AA) measured by powder 
x-ray diffraction for the as-cast samples of transition metal-doped
YCo$_5$ (green, crosses) and GdCo$_5$ (blue, stars).
\label{fig.lattice}
}
\end{figure}

In Fig.~\ref{fig.lattice} we show the lattice constants
$a$ and $c$ of the synthesized (polycrystalline) compounds 
measured by powder x-ray diffaction at room temperature.
\section{Computational details}
\label{app.compdet}

Our calculations proceed in two steps.
First, a self-consistent, scalar-relativistic calculation 
is performed on the magnetically-ordered system in order to
determine the potentials associated with each atomic species
(note that compositionally-disordered systems can be treated
at this step with the CPA).
Then, these potentials are fed into a non-self-consistent, 
fully-relativistic  CPA calculation
to model the magnetically-disordered system whose local moments
are orientated according to the probability distribution
specified by $\{\vec{\lambda_i}\}$.

For the first step,
we use the local-spin-density approximation for 
the exchange-correlation potential,\cite{Vosko1980}
treating the $4f$ electrons of Gd with the 
local-self-interaction correction.\cite{Lueders2005}
The Kohn-Sham potential is determined under a spherical
approximation, namely the atomic-sphere approximation (ASA).
The ASA sphere radii at the three distinct crystal sites
(RE, TM$_{2c}$, TM$_{3g}$) were (1.84, 1.39, 1.42)~\AA \ for YTM$_5$
and (1.85, 1.39, 1.42)~\AA \ for GdTM$_5$.
These values were chosen based on the results of a test
calculation performed on YCo$_5$ with the plane-wave projected-augmented
wave code \texttt{GPAW},\cite{Enkovaara2010} observing the radii at which the 
potentials centered at the three
sites showed similar deviations from spherical symmetry subject
to the ASA total volume constraint.

We investigated the spherical approximation further by performing 
calculations under the muffin-tin (MT) approximation
for the potential, which prohibits the overlap of different potential
spheres and consequently introduces a flat-potential interstitial region.
Our calculated critical temperatures based on MT calculations are
generally smaller than the ASA ones by $\sim$100~K, but trends (e.g.\
the relative critical temperatures of GdCo$_5$ and YCo$_5$, and
the effect of doping on different sites) are preserved.
However,  the calculated molecular field at the Gd site
is smaller in the MT approximation by almost a factor of 2.
Test calculations on the magnetocrystalline anisotropy 
also find that the MT approximation fails to predict the experimentally-observed
easy $c$-axis, while the ASA does.\cite{Nordstrom1992,Daalderop1996,Matsumoto2014}

These scalar-relativistic calculations are performed using
the \texttt{Hutsepot} KKR-CPA code.\cite{Daene2009}
The scattering matrices, Green's function etc.\ are 
expanded in a basis of spherical harmonics up to a maximum
angular momentum quantum number of $l=3$.
Although the KKR-CPA is an all-electron method, there
is still a partitioning of electrons into core and valence which
determines their treatment within multiple-scattering theory;
here the 4$p$ (5$p$) states were treated as valence
for Y (Gd).
A 20$\times$20$\times$20 Brillouin zone sampling was used
and a fixed electronic temperature of 400~K in calculating
the electronic occupations in the self-consistent calculation.

For the second step in our two-step procedure we solve the fully-relativistic
scattering problem\cite{Strange1984,Strange1989} using the 
previously-generated ``frozen'' potentials.
Here the $k$-space integration is performed to high
accuracy using an adaptive sampling algorithm.\cite{Bruno1997}
The electronic states were populated according to the Fermi-Dirac 
distribution whose temperature was chosen to match the local
moment statistics for $T \geq 400K$ and kept at 300~K otherwise.
The integration over angular variables 
in equation~\ref{eq.hint} was performed
numerically on a 240$\times$40 mesh equally spaced in $\sin\theta_i$ and
$\phi_i$, and the necessary energy integrations were performed
on a rectangular grid extending 2 Rydbergs into the complex plane,
using a logarithmic spacing with ten points per decade for the
legs of the contour parallel to the imaginary axis.
We note that the calculated electronic density could then
be used to construct new potentials in an iterative scheme,\cite{Deak2014} 
but here we keep the potentials frozen in line with 
the local moment picture.

Since the second part of the calculations is fully-relativistic,
the thermally-averaged
orbital angular momentum $\langle \mu_\mathrm{orb} \rangle_{0,T}$ 
can develop a nonzero value.
However,  the frozen potentials do not contain any explicit 
coupling to orbital
angular momentum, e.g.\ through an empirical orbital polarization correction
(OPC) term.\cite{Eriksson1990}
It has been observed that including such a term increases the
magnitude of the orbital moments in YCo$_5$ and also of the 
anisotropy.\cite{Daalderop1996,Nordstrom1992,Steinbeck20012}
Due to its empirical nature and the fact that it is largely 
untested for magnetically-disordered systems, we choose not to 
include an OPC term in the current study.

As mentioned in section~\ref{sec.theory}, the Weiss fields appear on both
sides of equation~\ref{eq.hint}, since
the $\{\vec{\lambda_i}\}$ values determine $P_0$.
Following Ref.~\citenum{Matsumoto2014} we obtain the Weiss fields
iteratively.
For lower temperatures $(\lambda \gtrsim 2)$ we find an
approach based on fixing $T$ to be efficient, i.e.\ the
$\lambda$-values for the next calculation are obtained
from the Weiss fields of the previous (prev) calculation as
\begin{equation}
\lambda_i^\mathrm{next} = \beta h_i^\mathrm{prev}
\end{equation}
for each sublattice $i$.
For smaller $\lambda$-values we find it more efficient to fix
$\lambda$; i.e.\ for sublattice $i$ $\lambda_i$ is fixed to some value (2, 1, 0.5, 0.1)
and $\lambda_j$ updated until a consistent solution is reached:
\begin{equation}
\lambda_j^\mathrm{next} = \lambda_i
\frac{h_j^\mathrm{prev}}{h_i^\mathrm{prev}}.
\end{equation}
Finally we note that we have a choice of magnetization direction
through the orientations of $\{\vec{\lambda_i}\}$.
To make contact with previous work\cite{Matsumoto2014} we
kept the magnetization direction fixed along [101]
and obtain the $h_i$ magnitudes for the iterative scheme 
by projecting onto the input $\vec{\lambda_i}$ direction.
We leave the important questions of magnetocrystalline anisotropy,
anisotropic exchange and magnetization anisotropy\cite{Gerard1992,Radwanski1992,
Alameda1981} for future study.

%\bibliography{papers}

\end{document}